\documentclass[12pt]{article}
\usepackage{amsmath}
\usepackage{amssymb}
\usepackage{epsfig}
\usepackage{graphicx}
\usepackage{cite}

\newcommand{\ba}{\begin{array}}
\newcommand{\ea}{\end{array}}
\newcommand{\Be}{\begin{equation}}
\newcommand{\Ee}{\end{equation}}

\newcommand{\bea}{\begin{eqnarray}}
\newcommand{\eea}{\end{eqnarray}}
\newcommand{\beas}{\begin{eqnarray*}}
\newcommand{\eeas}{\end{eqnarray*}}

\newcommand{\al}{\alpha}
\newcommand{\be}{\beta}
\newcommand{\Sol}  {\textrm{sol}}

\newcommand{\Atm}  {\textrm{atm}}

\newcommand{\unity}{1\hspace{-0.15cm}1}

\begin{document}

\vspace*{-1cm}
\phantom{hep-ph/***}
\hfill{IFIC/09-67}
\vskip 2.5cm
\begin{center}
{\Large\bf An $S_4$ model for quarks and leptons with maximal atmospheric angle \\
\vskip .3cm}
\end{center}
\vskip 0.5cm
\begin{center}
{\large Stefano Morisi}~$^{a)}$\footnote{e-mail address: morisi@ific.uv.es}
 and 
{\large Eduardo Peinado}~$^{a)}$\footnote{e-mail address: epeinado@ific.uv.es}
\\
\vskip .2cm
$^{a)}$~AHEP Group, Institut de F\'{\i}sica Corpuscular --
  C.S.I.C./Universitat de Val{\`e}ncia \\
  Edificio Institutos de Paterna, Apt 22085, E--46071 Valencia, Spain
\vskip 0.7cm
\end{center}

\begin{abstract}
We consider a model for quark and lepton masses and mixings based on $S_4$ flavor symmetry. The model contains six Higgs doublets where three of them give mass to the leptons and the other three gives mass to the quarks. Charged fermion and quark masses arise from renormalizable interactions while  
neutrino Majorana masses are generated through effective dimension five Weinberg operator. From the study of the minimization of the scalar potential
 we found a residual $\mu \leftrightarrow  \tau $ symmetry in the neutrino sector predicting
zero reactor angle and maximal atmospheric angle and for the quark sector we found a four-zero texture. We give a fit of the mass hierarchies and mixing angles in the quark sector.

\end{abstract}
\setcounter{footnote}{0}
\vskip2truecm

\section{Introduction}

Quarks and leptons have very different mixing angles.
A successful phenomenological ansatz for leptons has been proposed by
Harrison, Perkins and Scott and is given
by~\cite{Harrison:2002er}
\begin{equation}
\label{eq:HPS}
U_{\textrm{TBM}} = 
\left(\begin{array}{ccc}
\sqrt{2/3} & 1/\sqrt{3} & 0\\
-1/\sqrt{6} & 1/\sqrt{3} & -1/\sqrt{2}\\
-1/\sqrt{6} & 1/\sqrt{3} & 1/\sqrt{2}
\end{array}\right)
\end{equation}
which corresponds to 
$\tan^2\theta_{\Atm}=1$, $\sin^2\theta_{\textrm{Chooz}}=0$ and $\tan^2\theta_{\Sol}=0.5$, 
providing a good first approximation to the values indicated by
current neutrino oscillation data \cite{Schwetz:2008er,Fogli:2005cq}.
The third massive eigenstate is maximally mixed between $\mu$ and $\tau$ states
and the second eigenstate is trimaximally mixed  between $e,\,\mu$ and $\tau$.
Therefore the mixing matrix in eq.\,(\ref{eq:HPS}) is called 
{\it tri-bimaximal} (TBM) mixing matrix. While the experimental mixing matrix 
for  quarks is given by, see \cite{Charles:2004jd} 
\Be\label{ckmexp}
V_{\text{CKM}}=\left(
\begin{array}{ccc}
 0.9743 & 0.2252 & 0.0035 \\
 0.2251 & 0.97347 & 0.0412 \\
 0.00859 & 0.0404 & 0.999146
\end{array}
\right)
\Ee
In spite of the experimental progress so far we
have no a compelling theoretical evidence regarding the flavor problem,
namely why we have mixings \,(\ref{eq:HPS}) and (\ref{ckmexp})
and why fermions masses are hierarchical. 

A possibility to solve the flavor problem is by
assuming a symmetry between the three generations,  extending the standard model with a flavor symmetry $G_f$.
In past, when neutrino data was lacking, hypothesis on $G_f$ could arise
only from the quark sector. Successful ansatz for quarks was extended to the 
lepton sector, see for instance \cite{Pakvasa:1977in}. 
However recent discovery of large neutrino mixings suggest a different scenario.
Successfully ansatz for the lepton sector 
can be extended to the quark sector. Tri-bimaximal lepton mixing 
can be simply derived by assuming $A_4$ flavor symmetry\footnote{$A_4$
is the group of even permutations of four objects isomorphic to the group of symmetries of the tetrahedron.} \cite{TBA4} and other discrete flavor symmetries give nearly tri-bimaximal mixing~\cite{Mondragon:2007jx,Plentinger:2008up}. Example of extension with $A_4$ 
for the quark sector can be
found in \cite{He:2006dk,Bazzocchi:2007na,Lavoura:2007dw}.
In this paper we consider the lepton and quark sectors simultaneously.
There are two non Abelian discrete groups that are suitable for such a purpose, 
namely  $T'$\,\cite{Feruglio:2007uu,Chen:2007afa} and $S_4$~\cite{s4} since they contain singlet, doublet and triplet irreducible representations. It seems reasonable to consider models where
quarks transform as $2+1$ of $G_f$ and leptons as $3$ of $G_f$
in order to obtain large mixing in the lepton sector and small mixing
between first and second families in the quark sector with heavy top quark mass.
 
The  group of permutation of four objects $S_4$ is the minimal flavor symmetry of the mass matrix $M_l$
and $M_\nu$ yielding TBM as shown in \cite{Lam:2008sh}. However was recently clarified in \cite{Grimus:2009pg} that the 
symmetries of $M_l$ and $M_\nu$ are not also symmetries of the Lagrangian and thus $S_4$ is not special for TBM
but it is simply one of many groups that can be used for TBM.
Pioneer works using the symmetry group $S_4$ as a family symmetry and deducing predictions
for masses and mixings of  fermions  are in Ref.\,\cite{Pakvasa:1978tx}.
An interesting feature of $S_4$ is that the neutrino mass matrix generated from a general
dimension five Weinberg operator \cite{Weinberg:1979sa} like $LL\phi \phi/\Lambda$
where $\Lambda$ is the cut-off scale, 
is diagonalized from TBM when $\phi$ is an $A_4$-triplet $\phi=(\phi_1,\phi_2,\phi_3)$ that takes vev as 
$\langle \phi \rangle \sim (1,1,1)$. 
Differently in models with $A_4$
, the general dimension five operator $LL\phi \phi/\Lambda$ is
not diagonalized from TBM, see \cite{Morisi:2009sz}. The motivation is that in $A_4$ the contractions $(LL)_{1'}$ and
$(LL)_{1''}$ break TBM, while in $S_4$ the TBM is preserved since the two $A_4$ representations $1'$ and $1''$
correspond to a one irreducible representation of $S_4$, that is the doublet, and the contraction $(LL)_2$
preserve the TBM. 

In this paper, we study a model based on the $S_4$ flavor symmetry, where the model is invariant under the $G_F=S_4\times Z_3^q \times Z_2^q\rtimes (Z_{2e}\times Z_{2\mu} \times Z_{2\tau})$ product. The quarks and charged lepton masses arise from renormalizable interactions while the Majorana neutrino masses arise from the general dimension five operator. We also study the Higgs potential invariant under the $G_F$ symmetry and found the minimization conditions.

\vskip0.2cm

In the next section we introduce the model, in section 3
we study the minimization of the potential,
in sections 4 and 5 we study the phenomenological
consequence of our model for leptons and quarks respectively, and in section 6 we give the conclusions.

\section{The Model}
The model consist on the flavor symmetry $G_F=S_4\times Z_3^q \times Z_2^q\rtimes (Z_{2e}\times Z_{2\mu} \times Z_{2\tau})$, but in order to avoid unnecessary confusions, we will split the treatment for quarks and leptons. 

Consider the model defined in Table\,(\ref{tab1}).
\begin{table}[t]
\begin{center}
\begin{tabular}{|c|c|c|c|c|c|c|c|c|c|c|c|c|c|c|}
\hline
&$\overline{L}$&$e_R$&$\mu_R$&$\tau_R$ & $\overline{Q}_D$& $\overline{Q}_S$ & $u_{R_D}$&$u_{R_S}$ & $d_{R_D}$&$d_{R_S}$   &$\phi$ & $H_D$ & $H_s$\\
\hline
$SU(2)$&$2$&$1$&$1$&$1$ &2&2&1&1&1&1 &$2$&$2$&$2$\\\hline
$S_4$&$3_1$&$1_1$&$1_1$&$1_1$&$2$&$1$ &$2$&$1$   &$2$&$1$ & $3_1$& 2 & $1_1$\\\hline
$Z_3^q$& 1&1&1&1&$\omega$&$\omega$&$\omega^2$  & $\omega^2$&$\omega$  & $\omega$ &1&  $\omega$    & $\omega$\\
\hline
$Z_2^q$& $+$&$+$&$+$&$+$&$+$ &$-$&$+$ &$-$&$+$ &$-$     & $+$  &$-$&   $+$ \\
\hline
\end{tabular}
\caption{Quark, Lepton and scalar multiplet structure of our 
model, see text. }\label{tab1}
\end{center}
\end{table}
The left-handed doublets transform as a triplet of $S_4$, namely $L=(L_e,L_\mu,L_\tau)\sim 3_1$ and the right-handed 
fields $e_R$, $\mu_R$, $\tau_R$ as singlets of $S_4$. In the quark sector we assume the third family to transform
as a singlet of $S_4$ and the first and second families as a doublet, $Q_D=(Q_1,Q_2)$ and $q^c_D=(q_{R_1},q_{R_2})$. 
We have six Standard Model Higgs doublets,
$\phi=(\phi_1,\phi_2,\phi_3)$ transforming as a triplet under $S_4$, $H_D=(H_1,H_2)$ and $ H_s$ transforming 
as a doublet and a singlet respectively under the $S_4$ flavor symmetry. 
Only the quark sector is charged with respect to the $Z_3^q$ and $Z_2^q$ 
symmetries. 
As a consequence of $S_4\times Z_3^q\times Z_2^q$ assignment, the scalar field $\phi$ interact only with charged leptons
and $H_{D,S}$ only with the quarks at the renormalizable level.

In order to have diagonal charged lepton mass matrix we also assume extra auxiliary
symmetries $Z_{2e}\times Z_{2\mu} \times Z_{2\tau}$. 
In particular each right-handed field $l^c_a$ is charged under the corresponding $Z_{2a}$ with $a=e,\mu,\tau$ 
as well as each component of $\phi$. 
As in \cite{Grimus:2005rf,Mohapatra:2006pu,Morisi:2009qa} the $(Z_2)^3$ symmetries glue each $l_i^c$ with the corresponding $\phi_i$. 
We will show below that  $(Z_2)^3$  remove off-diagonal terms in the charged lepton sector. Since the  $(Z_2)^3$
symmetries do not commute with $S_4$ we have to take the semidirect product\footnote{For the use of semi-direct product in model building see for instance~\cite{Morisi:2009qa,semi}.} of
$S_4$ with $ (Z_{2e}\times Z_{2\mu} \times Z_{2\tau})$. The assignment of the charged leptons and the scalar field $\phi$
with respect to $S_4\rtimes (Z_{2e}\times Z_{2\mu} \times Z_{2\tau})$ can be summarized in the following table.
\begin{center}
\begin{tabular}{|c|c|c|c|c||c|c|c|c|c|c|}
\hline
leptons &$L_{}$&$e^c$&$\mu^c$&$\tau^c$   &$\phi_1$ & $\phi_2$ & $\phi_3$\\
\hline
$S_4$&$3_1$&$1_1$&$1_1$&$1_1$&\multicolumn{3}{c|}{$3_1$}\\\hline
$Z_{2e}$   &$+$&$-$&$+$&$+$&$-$&$+$&$+$\\\hline
$Z_{2\mu}$ &$+$&$+$&$-$&$+$&$+$&$+$&$-$\\\hline
$Z_{2\tau}$&$+$&$+$&$+$&$-$&$+$&$-$&$+$\\\hline
\end{tabular}
\end{center}
%
%
%
The $S_4\times Z_3^q \times Z_2^q\rtimes (Z_{2e}\times Z_{2\mu} \times Z_{2\tau})$ invariant Lagrangian reads
$$
\mathcal{L}=\mathcal{L}_l+\mathcal{L}_M+\mathcal{L}_q
$$
where
\begin{eqnarray}
\label{eq:LL}
\mathcal{L}_l&=&{y_{e}}\overline{L}_e e_R\phi_1+y_\mu \overline{L}_\mu  \mu_R\phi_3+y_\tau \overline{L}_\tau  \tau_R\phi_2\,;\\
&&\nonumber\\
\mathcal{L}_{M}&=&
\frac{\lambda_1}{\Lambda} (L_i\tilde{\phi})_1 (L_j\tilde{\phi})_1+
\frac{\lambda_2}{\Lambda} (L_i\tilde{\phi})_{2} (L_j\tilde{\phi})_{2}+
\frac{\lambda_3}{\Lambda} (L_i\tilde{\phi})_{3_1} (L_j\tilde{\phi})_{3_1}+
\frac{\lambda_3'}{\Lambda} (L_i\tilde{\phi})_{3_2} (L_j\tilde{\phi})_{3_2}+\nonumber\\
&+&\frac{\lambda_4}{\Lambda} (L_iL_j)_1 (\tilde{\phi} \tilde{\phi})_1+
\frac{\lambda_5}{\Lambda} (L_iL_j)_{2} (\tilde{\phi} \tilde{\phi})_{2}+
\frac{\lambda_6}{\Lambda} (L_iL_j)_{3_1} (\tilde{\phi} \tilde{\phi})_{3_1}
\,;\label{S4}\\
&&\nonumber\\
\mathcal{L}_{Y}^q&=&
Y^d_2  \overline{Q}_D d_{R_D} H_s+Y^d_3 \overline{Q}_S d_{R_S} H_s+Y^d_4 \overline{Q}_D d_{R_S} H_D+Y^d_5 \overline{Q}_S q_{R_D} H_D+\nonumber\\
&+&Y^u_2  \overline{Q}_D u_{R_D} \tilde{H}_s+Y^u_3 \overline{Q}_S u_{R_S} \tilde{H}_s+Y^u_4 \overline{Q}_D u_{R_S} \tilde{H}_D+Y^u_5 \overline{Q}_S u_{R_D} \tilde{H}_D,
\label{lquark}
\end{eqnarray}
where $\Lambda$ is a cut-off scale and $\tilde{\phi}=i \sigma_2 \phi^*$ and so on.
According to the $S_4$ symmetry also the following off-diagonal terms in the lepton sectors are allowed
\begin{eqnarray}
&&y_{e}  (\overline{L}_\mu  e_R \phi_3 + \overline{L}_\tau  e_R\phi_2 )+
y_\mu  (\overline{L}_e \mu_R\phi_1+\overline{L}_\tau \mu_R\phi_2)+
y_\tau (\overline{L}_e \tau_R\phi_1+\overline{L}_\mu \tau_R\phi_3),
%
\end{eqnarray}
but these terms are not invariant under the $Z_{2e}\times Z_{2\mu} \times Z_{2\tau}$ symmetry.

\section{The scalar potential}

In our model there are 6 Higgs doublets that belong to one triplet $\phi=(\phi_1,\phi_2,\phi_3)$, one doublet 
$H_D=(H_1,H_2)$ and one singlet $H_s$ 
representations of $S_4$. 
In general the Higgs potential can be written as
\Be\label{eq:higgs}
V=V(\phi)+V(H_D,H_s)+V_{\text{int}}(\phi,H_D,H_s)
\Ee
where $V(\phi)$ contains only $S_4$-triplet, $V(H_D,H_s)$ contains both $S_4$ singlet and doublet scalars 
and $V_{\text{int}}(\phi,H_D,H_s)$ contains only quartic terms mixing the triplet with the doublet and the singlet.
Moreover $H_D$ and $H_s$ transform with
respect to $Z_3^q\times Z_2^q$ and each component $\phi_i$ of the $S_4$ triplet
$\phi$ transforms with respect to $Z_{2e}\times Z_{2\mu} \times Z_{2\tau}$. Below
we give  the potential invariant under $S_4 \times Z_3^q \times Z_2^q$ and successively 
we consider  the $Z_{2e}\times Z_{2\mu} \times Z_{2\tau} $ symmetry  when  we 
explicitly write each term in its components.  The contributions to the Higgs Potential in eq. (\ref{eq:higgs}) invariant under the $S_4 \times Z_3^q\times Z_2^q$ symmetry are given by\footnote{The term proportional to $l_6$ is of the form $(H^{\dagger}H_s)^2$}
\Be
\ba{rcl}
V(\phi)&=&\mu [\phi^{\dagger}\phi]_{1_1}+\alpha ([\phi^{\dagger}\phi]_{1_1})^2+l_2 [\phi^{\dagger}\phi]_2[\phi^{\dagger}\phi]_2+
l_3[\phi^{\dagger}\phi]_{3_1}[\phi^{\dagger}\phi]_{3_1}+\\ \\&+&l_4[\phi^{\dagger}\phi]_{3_2}[\phi^{\dagger}\phi]_{3_2},\\ \\
V(H_D,H_s)&=&\mu_2 [H_D^{\dagger}H_{D}]_{1_1}+\mu_3[H_s^{\dagger}H_s]+\alpha_2([H_D^{\dagger}H_{D}]_{1_1})^2+\alpha_3 ([H_D^{\dagger}H_{D}]_{1_2})^2+\\ \\ &+&\alpha_4[H_D^{\dagger}H_{D}]_{2}[H_D^{\dagger}H_{D}]_{2}+l_5[H_D^{\dagger}H_{D}]_{1_1}[H_s^{\dagger}H_{s}]_{1_1}+\alpha_5[H_s^{\dagger}H_{s}]^2+\\ \\&+&  l_6 [H_D^{\dagger}H_{D}^{\dagger}]_{1_1}[H_sH_{s}]_{1_1} ,\\ \\
V_{\text{int}}(\phi,H_D,H_s)&=&a_1[\phi^{\dagger}\phi]_{1_1}[H_s^{\dagger}H_{s}]_{1_1}+b_1[\phi^{\dagger}\phi]_{1_1}[H_D^{\dagger}H_{D}]_{1_1}+ b_2[\phi^{\dagger}\phi]_{2}[H_D^{\dagger}H_{D}]_{2}\\ \\ &+&c_3[H_{D}^{\dagger}\phi]_{3_1}[\phi^{\dagger}H_{D}]_{3_1}+c_4[H_{D}^{\dagger}\phi]_{3_2}[\phi^{\dagger}H_{D}]_{3_2}
\ea
\Ee
Once the symmetry $Z_{2e}\times Z_{2\mu} \times Z_{2\tau}$ is imposed, the contribution to the potential coming from $V_{\text{int}}(\phi,H_D,H_s)$ is reduced to
\Be
V_{\text{int}}(\phi,H_D,H_s)=\left[(b_1+c_3-c_4) (H_1^{\dagger} H_1+H_2^{\dagger} H_2)+a_1H_s^{\dagger} H_s\right](\phi_1^{\dagger}\phi_1+\phi_2^{\dagger}\phi_2+\phi_3^{\dagger}\phi_3),
\Ee
where we can reabsorb $c_3$ and $c_4$ in $b_1$. Similarly we reduce the other terms after considering the symmetry $Z_{2e}\times Z_{2\mu} \times Z_{2\tau}$, then 
the full scalar potential $V$ invariant under   $S_4\times Z_3^q\times Z_2^q \rtimes (Z_{2e}\times Z_{2\mu} \times Z_{2\tau})$
is given by
\Be
\ba{rll}
V&=& \mu(\phi_1 \phi_1^{\dagger}+\phi_2 \phi_2^{\dagger}+\phi_3 \phi_3^{\dagger}) +(4 l_3+\alpha ) (\phi_1 \phi_1^{\dagger})^2+
\\&+&2 (l_2+5 l_3-l_4+\alpha ) \phi_2\phi_2^{\dagger}\phi_3\phi_3^{\dagger}+(l_3+l_4+\alpha ) ((\phi_2 \phi_2^{\dagger})^2+(\phi_3 \phi_3^{\dagger})^2)+\\&+&2 (l_2-l_3+l_4+\alpha ) \phi_1 \phi_1^{\dagger} (\phi_3 \phi_3^{\dagger}+\phi_2 \phi_2^{\dagger})+\\
&&\\
&+&\left(H_1H_1^{\dagger}+H_2H_2^{\dagger}\right)^2\alpha_2+\left(-H_1H_1^{\dagger}+H_2H_2^{\dagger}\right)^2 \alpha_3+2H_1H_1^{\dagger} H_2H_2^{\dagger} \alpha_4+\\ &+&\left(H_1H_1^{\dagger}+H_2H_2^{\dagger}\right) \mu_2+\left(H_1H_1^{\dagger}+H_2H_2^{\dagger}\right) l_5 H_s H_s^{\dagger}+\\ &+&\mu_3 H_s H_s^{\dagger}+\alpha_5 (H_s H_s^{\dagger})^2+2l_6 H_1^{\dagger}H_2^{\dagger}H_sH_s+h.c.\\
&&\\
&+&b_1(H_1^{\dagger}H_1+H_2^{\dagger}H_2)(\phi_1 \phi_1^{\dagger}+\phi_2 \phi_2^{\dagger}+\phi_3 \phi_3^{\dagger})+\\ &+&a_1 H_s^{\dagger}H_s(\phi_1 \phi_1^{\dagger}+\phi_2 \phi_2^{\dagger}+\phi_3 \phi_3^{\dagger})\label{pot}
\ea
\Ee
In the case of real vev's, that is
\Be
\ba{l}
\langle \phi \rangle = (v_1,~v_2,~v_3)\\
\\
\langle H_D \rangle = (h_1,~h_2)\\
\\
\langle H_s \rangle=v_s
\ea
\Ee 
%
the equations of minimum are
\Be
\ba{lcl}
\frac{\partial V}{\partial v_1}&=&2 v_1 \left(b_1 \left(h_1^2+h_2^2\right)+l_3 \left(8 v_1^2-2 \left(v_2^2+v_3^2\right)\right)+a_1 v_s^2+2 v_1^2 \alpha +2 \left(v_2^2+v_3^2\right) (l_2+l_4+\alpha )+\mu \right)
\\ \\&=&0,
\\ \\
\frac{\partial V}{\partial v_2}&=&
2 v_2 \left[b_1 \left(h_1^2+h_2^2\right)+a_1 v_s^2+\mu+ \right.\\ \\
&+&
\left.2 \left(l_4 \left(v_1^2+v_2^2-v_3^2\right)+l_2 \left(v_1^2+v_3^2\right)+  l_3 \left(-v_1^2+v_2^2+5 v_3^2\right)+\left(v_1^2+v_2^2+v_3^2\right) \alpha \right) \right]
\\ \\&=&0,
\\ \\
\frac{\partial V}{\partial v_3}&=&
2 v_3 \left[b_1 \left(h_1^2+h_2^2\right)+a_1 v_s^2+\mu\right.\\ \\
&+& 
\left.2 \left(l_2 \left(v_1^2+v_2^2\right)+l_4 \left(v_1^2-v_2^2+v_3^2\right)+  l_3 \left(-v_1^2+5 v_2^2+v_3^2\right)+\left(v_1^2+v_2^2+v_3^2\right) \alpha \right)+ \right]
\\ \\&=&0,
\ea
\Ee
and
\Be
\ba{lcl}
\frac{\partial V}{\partial h_1}&=&
2 h_1 \left(b_1 \left(v_1^2+v_2^2+v_3^2\right)+l_5 v_s^2+2 h_1^2 (\alpha_2+\alpha_3)+2 h_2^2 (\alpha_2-\alpha_3+\alpha_4)+\mu_2\right)+2l_6h_2v_s^2=0,
\\ \\
\frac{\partial V}{\partial h_2}&=&
2 h_2 \left(b_1 \left(v_1^2+v_2^2+v_3^2\right)+l_5 v_s^2+2 h_2^2 (\alpha_2+\alpha_3)+2 h_1^2 (\alpha_2-\alpha_3+\alpha_4)+\mu_2\right)+2l_6h_1v_s^2=0,
\\ \\
\frac{\partial V}{\partial v_s}&=&
2 v_s \left(\left(h_1^2+h_2^2\right) l_5+a_1 \left(v_1^2+v_2^2+v_3^2\right)+2 v_s^2 \alpha_5\mu_3+2l_6h_1h_2\right)=0.
\label{vmin}
\ea
\Ee
From the second and third equations we obtain $v_2=v_3\equiv v$. From the first and second equation we found 
\Be
v_1=\frac{\sqrt{l_2-8 l_3+2 l_4} v_3}{\sqrt{l_2-5 l_3+l_4}}\equiv r v
\Ee
We redefine $h_2=\delta h_1$, and the remaining equations are rewritten as
\Be
\ba{l}
2 v^2 \left(2 \left(l_2-l_3+\text{l4}+2 l_3 r^2\right)+\left(2+r^2\right) \alpha \right)+\text{b1} h_1^2 \left(1+\delta ^2\right)+\mu +a_1 \chi ^2=0
\\ \\
\text{b1} \left(2+r^2\right) v^2+2 h_1^2 \left(\alpha_2+\alpha_3+(\alpha_2-\alpha_3+\alpha_4) \delta ^2\right)+\mu_2+(l_5+l_6 \delta ) \chi ^2=0
\\ \\
  \left(\text{b1} \left(2+r^2\right) v^2+2 h_1^2 \left(\alpha_2-\alpha_3+\alpha_4+(\alpha_2+\alpha_3) \delta ^2\right)+\mu_2\right)+\frac{(l_6+l_5 \delta ) \chi ^2}{\delta }=0
\\ \\
a_1 \left(2+r^2\right) v^2+h_1^2 \left(l_5+2 l_6 \delta +l_5 \delta ^2\right)+\mu_3+2 \alpha_5 \chi ^2b_1 h_1^2+a_1 v_s^2+\\ \\+2 v_3^2 \left(l_2 \left(1+r^2\right)+2 (3 l_3+\alpha )+r^2 (-l_3+l_4+\alpha )\right)+\mu =0
\ea
\Ee
From the second and third of these equations we found two solutions for $\delta$, one is $\delta=\pm 1$, and the other is the one we are interested in
\Be
\delta =\frac{l_6 \chi ^2}{2 h_1^2 (2 \alpha_3-\alpha4)},
\Ee
it is clear that in the limit $l_6\rightarrow 0$, then $\delta\rightarrow 0$.

From the rest of equations we can determine $v$, $v_s$ and $h_1$.
It is straightforward to compute the Hessian $\partial^2 V/ \partial u_i \partial u_j$, where $u=(v_1,v_2,v_3,h_1,h_2,v_s)$, of the Higgs potential. 
 We found that for a large region of the parameter space the Hessian is definite
positive, therefore the solution we found is a real minimum.
Summarizing the structure of the vev's, 
for the doublet and the singlet of $S_4$ we have $H_1=h_1$, $H_2=\delta h_1$ and $H_s=v_s$, so the alignment of the doublet is of the form
\begin{equation}
H_D\sim(1,\delta).
\end{equation}
In the limit $l_6=0$, the alignment takes the form $H_D\sim(1,0).$

For the triplet of $S_4$, $v_1=rv$ and $v_2=v_3\equiv v$, so the alignment is of the form 
\begin{equation}\label{allphi}
\phi\sim(r,1,1).
\end{equation}
Notice that in the case $l_4=3l_3$, implies $r=1$, and the alignment takes the form $\phi\sim(1,1,1).$
\section{Lepton sector and maximal atmospheric angle}


The charged lepton mass matrix is diagonal with masses proportional to the Yukawa couplings $y_e, \, y_\mu$
and $y_\tau$.


When the  $\phi$ Higgs doublet takes vev as, see equation\,(\ref{allphi}), 
\[
\phi\sim(r,1,1),
\]
the Majorana neutrino mass matrix takes the form
\begin{eqnarray}
M_\nu&=&
\l_1
\left(\begin{array}{ccc}
r^2 &r &r \\
 &1 &1 \\
 & &1 \\
\end{array}\right)+
\l_2
\left(\begin{array}{ccc}
2 &1+r &1+r \\
 &2r &1+r^2 \\
 & &2r \\
\end{array}\right)+
\l_3
\left(\begin{array}{ccc}
2+4 r^2 &-2 -r&-2-r \\
 &1 -4r& 5+r^2\\
 & & 1-4r\\
\end{array}\right)+\nonumber\\
&+&\l_3'
\left(\begin{array}{ccc}
-2 &r &r \\
 & 1& -1-r^2\\
 & & 1\\
\end{array}\right)+
\l_4
\left(\begin{array}{ccc}
1 &0 &0 \\
 &0 &1 \\
 & &0 \\
\end{array}\right)+
\l_5
\left(\begin{array}{ccc}
0 & 1&1 \\
 &1 &0 \\
 & &1 \\
\end{array}\right)\\
&+&
\l_6
\left(\begin{array}{ccc}
2(2 r^2-2) &-2+2 r &-2+2 r \\
 & 2(2-2r)&-2r^2+2 \\
 & &2(2-2r) \\
\end{array}\right)\equiv
\left(\begin{array}{ccc}
x &y &y \\
y &z &w \\
y &w &z \\
\end{array}\right)
\nonumber
\end{eqnarray}
where $\l_i=v^2 \lambda_i/\Lambda$ and $r=v_1/v_3$.
The matrix $M_\nu$ is $\mu\leftrightarrow\tau$ invariant 
therefore the atmospheric angle is maximal and the reactor angle is zero \cite{Grimus:2004rj}. 
The solar angle depend from the parameter $r$ and it is unpredicted. In the limit $r\to 1$ (see the potential) we have that
$x+y = w+z $ and the solar angle is trimaximal.

We have three different  eigenvalues:
\Be
\ba{l}
m_1=\frac{1}{2}\left(w+x+z-\sqrt{w^2+x^2+8y^2+z^2+2wz-2x(w+z)}\right)\\
\\
m_2=\frac{1}{2}\left(w+x+z+\sqrt{w^2+x^2+8y^2+z^2+2wz-2x(w+z)}\right)\\
\\
m_3=z-w,
\ea
\Ee
and it is possible to reproduce  the ratio 
$\alpha=\Delta m_{\textrm{sol}}^2/\Delta m_{\textrm{atm}}^2$.

\section{The quark sector}

In order to have small mixings and a hierarchical 
mass pattern for quarks we associate the quarks $Q_L$ and $q^{c}_{L}$ in a $2\oplus 1_1$ irreducible representation 
of $S_4$.
The assignment for particles of the model are shown in table~\ref{tab1}. 
From the Yukawa Lagrangian in eq.\,(\ref{lquark}), 
once the electroweak symmetry is broken we obtain the mass matrix for the quarks
\Be
M_q=\left(
\ba{ccc}
0 & Y_2 h_s & Y_4 h_2 \\
Y_2 h_s & 0 & Y_4 h_1 \\
Y_5 h_2 & Y_5 h_1 & Y_3 h_s 
\ea
\right).
\label{quarkmass}
\Ee
From the study of the scalar potential we have the alignment
\Be
\langle H_D \rangle\sim (1,\delta).
\Ee
As we know the nearest neighbor interaction (NNI) form of the quarks matrices is in agreement with the experiments of quark masses and mixings, so $\delta$ must be very small. This limit is obtained by setting $l_6=0$ in the Higgs potential. With this fine tunning, the alignment for $H_D$ is $H_D\sim (~1,~0)$ and the quark mass matrices, in eq. (\ref{quarkmass}) take the form
\Be
M_{u,d}=\left(
\ba{ccc}
0 & Y_2^{u,d} h_s & 0 \\
Y_2^{u,d} h_s & 0 & Y_4^{u,d} h_1 \\
0 & Y_5^{u,d} h_1 & Y_3^{u,d} h_s 
\ea
\right).\label{massquarks}
\Ee
Such matrices are four-zero texture and have the form of a nearest neighbor interaction, first proposed by Weinberg~\cite{Weinberg:1977hb} and then extended by Fritzsch~\cite{Fritzsch:1977vd} (see also~\cite{Babu:2004tn} and references therein). The mass matrices in (\ref{massquarks}) have factorizable phases~\cite{Branco:2004ya,Babu:2009fd}, i.e. $M_{u,d}=P^{u,d}_L\tilde{M}_{u,d}P^{u,d}_R$, where $\tilde{M}$ has the same structure as (\ref{massquarks}) but without phases. Only two combinations  of phases will enter into the CKM matrix, $P=P^{u\star}_{L}P^d_L=\mbox{diag}(1,e^{i\beta_{ud}},e^{i\alpha_{ud}})$.

We  rewritte the mass matrices $\tilde{M}_{u,d}$ as
\Be
\tilde{M}_{u,d}=m_{t,b}\left(
\ba{ccc}
0 & \frac{q_{u,d}}{y_{u,d}} & 0 \\
\frac{q_{u,d}}{y_{u,d}}& 0 & b_{u,d}  \\
0 & d_{u,d} &  y_{u,d}^2
\ea
\right).
\Ee
where 
\begin{eqnarray}
b_{u,d} & = & \sqrt{\frac{p_{u,d}+1-y_{u,d}^{4} - R_{u,d}}{2} ~ - ~ \frac{q_{u,d}^2}{y_{u,d}^2}} \; , \nonumber \\
d_{u,d} & = & \sqrt{\frac{p_{u,d}+1-y_{u,d}^{4} + R_{u,d}}{2} ~ - ~ \frac{q_{u,d}^2 }{y_{u,d}^2}} \; ,
\label{bd}
\end{eqnarray}
and
\begin{equation}
R_{u,d} \equiv \sqrt{\left (1+p_{u,d}-y_{u,d}^4\right )^2 - 4 \left (p_{u,d}+q_{u,d}^4\right )
+ 4 q_{u,d}^2 y_{u,d}^2 2}.
\label{R}
\end{equation}
where $p_{u,d}$ and $q_{u,d}$ defined by
\Be\ba{lr}
p_{u} =\frac{m_{u} m_{c}}{m^2_{t}}& p_{d} =\frac{m_{d} m_{s}}{m^2_{b}}\\ \\ 
q_{u}^2 = \frac{m_{u}^2+ m_{c}^2}{m^2_{t}} & q_{d}^2 = \frac{m_{d}^2+ m_{s}^2}{m^2_{b}}.
\ea\Ee

In this case, we have 4 real parameters in each mass matrix, $q_{u,d},~b_{u,d},~d_{u,d}$ and $y_{u,d}$. These parameters are rewritten in term of the masses of the quarks and the free parameters $y_{u,d}$. Therefore in the mixing matrix appears, 6 masses, two real parameters $y_{u,d}$ and the relative phases $\alpha_{ud}=\alpha_u-\alpha_d$ and 
$\beta_{ud}=\beta_u-\beta_d$.
The CKM matrix is then given by 
\Be
V_{CKM}=O_{u}^TPO_{d},
\Ee
where $O_{u,d}$ are the orthogonal matrices that diagonalize $\tilde{M}_{u,d}$ via
\Be
O_{u,d}^T\tilde{M}_{u,d}\tilde{M}_{u,d}^TO_{u,d}=\mbox{diag}(m_{u,d}^2,m_{c,s}^2,m_{t,b}^2).
\Ee

In this case we can fit the quark mixing angles with a very good precision,
with the values 
\begin{equation}
\ba{lll}
y_u= 0.996333 & y_d= 0.957981&  \\ 
\alpha_{ud} = -2.03052&\beta_{ud} =-1.49938 &\\
m_u= 2.35634~MeV& m_c= 1237.37~MeV& m_t= 174.276~GeV\\ 
m_d= 5.27743~MeV& m_s= 90.8056~MeV& m_b= 4243.63~MeV
\ea
\end{equation}
we found
\Be
V^{\text{th}}=\left(
\begin{array}{ccc}
 0.974328 & 0.2251 & 0.00380224 \\
 0.22497 & 0.973513 & 0.0407534 \\
 0.00855318 & 0.0400268 & 0.999162
\end{array}
\right)
\Ee
which is in excellent agreement with the experimental values in eq. (\ref{ckmexp}). The prediction for the Jarlskog invariant is $J=3.1\times 10^{-5}$ which is also in agreement withe the experimental central value $J^{exp}=(2.92\pm 0.15)\times 10^{-5} $.
Note that one of the phases in $P$ is redundant in the sense that to make a fit of quark mixings it is enough with one phase, as in \cite{Babu:2004tn}.

\section{Conclusions}
We have studied a model based on the flavor symmetry $S_4$ for leptons and
quarks where all charged fermion masses arise from renormalizable Lagrangian
and neutrino mass matrix is induced  from dimension five Weinberg operator.
The model contains six Standard Model Higgs fields transforming respectively 
as a triplet, a doublet and a singlet  under $S_4$. We have studied 
the minimization of the full potential.
We obtain a $\mu-\tau$ exchange invariant neutrino mass matrix predicting
maximal atmospheric angle while the solar angle is undetermined. The model is compatible with normal, inverse and degenerate hierarchies for the neutrino masses. 
The quark mass matrices pattern have the form of a nearest neighbor interaction with four-zero texture. 
We give a numerical solution 
showing that it is possible to  reproduce correctly
 the CKM mixing matrix within the experimental errors. It is well known that models with more than one Higgs $SU(2)$ doublet may in general, have tree level
flavor changing neutral currents (FCNC)~\cite{Glashow:1976nt}. A complete analysis of this problem as well as a more deep analysis of the Higgs phenomenology, masses and decays widths, will go beyond the scope of the present paper, and we would like to leave this problems to a future work. An analysis of these problems was done in the scenario of an $S_3$ flavor symmetry in~\cite{Kaneko:2006wi} where the strongest constraint on the scalar masses arises from the neutral K meson mixing.

\section*{Acknowledgments}

Work supported by the EC contract UNILHC PITN-GA-2009-237920,
by the Spanish grants FPA2008-00319 and CDS2009-00064 (MICINN)
and PROMETEO/2009/091 (Generalitat Valenciana)
and by European Commission Contracts
MRTN-CT-2004-503369 and ILIAS/N6 RII3-CT-2004-506222.

\appendix


\section{The group $S_4$}
$S_4$ is the finite group of the permutations of four objects (for a short
introduction to $S_4$, see for instance \cite{Hagedorn:2006ug} and references therein). 
$S_4$ has 5 irreducible representations, two singlets $1_1$ and $1_2$, a doublet 2,
and two triplets $3_1$ and $3_2$. 

The group $S_4$ is defined can be defined by  two generators  $S$ and $T$ that satisfy
\begin{equation}\label{rel}
 S^4= T^3=  (ST^2)^2=\unity \,.
 \end{equation}
In the basis of $T$ diagonal the generators can be written for the different representations as 
\begin{description}
  \item[representation $1_1$:] $S=1$, $T=1$
  \item[representation $1_2$:] $S=-1$, $T=1$
  \item[representation $2$:] $S=\left(
                               \begin{array}{cc}
                                 0 & 1 \\
                                 1 & 0 \\
                               \end{array}
                             \right)$, $T=\left(
                                            \begin{array}{cc}
                                              \omega & 0 \\
                                              0 & \omega^2 \\
                                            \end{array}
                                          \right)$
  \item[representation $3_1$:] $S=\dfrac{1}{3}\left(
                                 \begin{array}{ccc}
                                   -1 & 2\omega & 2\omega^2 \\
                                   2\omega & 2\omega^2 & -1 \\
                                   2\omega^2 & -1 & 2\omega \\
                                 \end{array}
                               \right)$, $T=\left(
                                              \begin{array}{ccc}
                                                1 & 0 & 0 \\
                                                0 & \omega^2 & 0 \\
                                                0 & 0 & \omega \\
                                              \end{array}
                                            \right)$
  \item[representation $3_2$:] $S=\dfrac{1}{3}\left(
                                 \begin{array}{ccc}
                                   1 & -2\omega & -2\omega^2 \\
                                   -2\omega & -2\omega^2 & 1 \\
                                   -2\omega^2 & 1 & -2\omega \\
                                 \end{array}
                               \right)$, $T=\left(
                                              \begin{array}{ccc}
                                                1 & 0 & 0 \\
                                                0 & \omega^2 & 0 \\
                                                0 & 0 & \omega \\
                                              \end{array}
                                            \right)$\;.
\end{description}
In this basis the rules for the non trivial products of two irreducible representations are:
\begin{description}
\item The multiplication rules with the 2-dimensional
representation are the following:
\[
\begin{array}{ll}
2\otimes2=1_1\oplus1_2\oplus2&\quad
\text{with}\quad\left\{\begin{array}{l}
                    1_1\sim\alpha_1\beta_2+\alpha_2\beta_1\\[-10pt]
                    \\[8pt]
                    1_2\sim\alpha_1\beta_2-\alpha_2\beta_1\\[-10pt]
                    \\[8pt]
                    2\sim\left(\begin{array}{c}
                        \alpha_2\beta_2 \\
                        \alpha_1\beta_1 \\
                    \end{array}\right)
                    \end{array}
            \right.\\[-10pt]
\\[8pt]
2\otimes3_1=3_1\oplus3_2&\quad
\text{with}\quad\left\{\begin{array}{l}
                    3_1\sim\left(\begin{array}{c}
                        \alpha_1\beta_2+\alpha_2\beta_3 \\
                        \alpha_1\beta_3+\alpha_2\beta_1 \\
                        \alpha_1\beta_1+\alpha_2\beta_2 \\
                    \end{array}\right)\\[-10pt]
                    \\[8pt]
                    3_2\sim\left(\begin{array}{c}
                        \alpha_1\beta_2-\alpha_2\beta_3\\
                        \alpha_1\beta_3-\alpha_2\beta_1 \\
                        \alpha_1\beta_1-\alpha_2\beta_2 \\
                    \end{array}\right)\\
                    \end{array}
            \right.\\[-10pt]
\\[8pt]
2\otimes3_2=3_1\oplus3_2&\quad
\text{with}\quad\left\{\begin{array}{l}
                    3_1\sim\left(\begin{array}{c}
                        \alpha_1\beta_2-\alpha_2\beta_3\\
                        \alpha_1\beta_3-\alpha_2\beta_1 \\
                        \alpha_1\beta_1-\alpha_2\beta_2 \\
                    \end{array}\right)\\[-10pt]
                    \\[8pt]
                    3_2\sim\left(\begin{array}{c}
                        \alpha_1\beta_2+\alpha_2\beta_3 \\
                        \alpha_1\beta_3+\alpha_2\beta_1 \\
                        \alpha_1\beta_1+\alpha_2\beta_2 \\
                    \end{array}\right)\\
                    \end{array}
            \right.\\
\end{array}
\]

\item The multiplication rules with the 3-dimensional
representations are the following:
\[
\begin{array}{ll}
3_1\otimes3_1=3_2\otimes3_2=1_1\oplus2\oplus3_1\oplus3_2\qquad
\text{with}\quad\left\{
\begin{array}{l}
1_1\sim\alpha_1\beta_1+\alpha_2\beta_3+\alpha_3\beta_2 \\[-10pt]
                    \\[8pt]
2\sim\left(
     \begin{array}{c}
       \al_2\be_2+\al_1\be_3+\al_3\be_1 \\
       \al_3\be_3+\al_1\be_2+\al_2\be_1 \\
     \end{array}
   \right)\\[-10pt]
   \\[8pt]
3_1\sim\left(\begin{array}{c}
         2\al_1\be_1-\alpha_2\beta_3-\alpha_3\beta_2 \\
         2\al_3\be_3-\alpha_1\beta_2-\alpha_2\beta_1 \\
         2\al_2\be_2-\alpha_1\beta_3-\alpha_3\beta_1 \\
        \end{array}\right)\\[-10pt]
        \\[8pt]
3_2\sim\left(\begin{array}{c}
         \alpha_2\beta_3-\alpha_3\beta_2 \\
         \alpha_1\beta_2-\alpha_2\beta_1 \\
         \alpha_3\beta_1-\alpha_1\beta_3 \\
    \end{array}\right)
\end{array}\right.
\end{array}
\]
\[
\begin{array}{ll}
3_1\otimes3_2=1_2\oplus2\oplus3_1\oplus3_2\qquad
\text{with}\quad\left\{
\begin{array}{l}
1_2\sim\alpha_1\beta_1+\alpha_2\beta_3+\alpha_3\beta_2\\[-10pt]
        \\[8pt]
2\sim\left(
     \begin{array}{c}
       \al_2\be_2+\al_1\be_3+\al_3\be_1 \\
       -\al_3\be_3-\al_1\be_2-\al_2\be_1 \\
     \end{array}
   \right)\\[-10pt]
        \\[8pt]
3_1\sim\left(\begin{array}{c}
         \alpha_2\beta_3-\alpha_3\beta_2 \\
         \alpha_1\beta_2-\alpha_2\beta_1 \\
         \alpha_3\beta_1-\alpha_1\beta_3 \\
    \end{array}\right)\\[-10pt]
        \\[8pt]
3_2\sim\left(\begin{array}{c}
         2\al_1\be_1-\alpha_2\beta_3-\alpha_3\beta_2 \\
         2\al_3\be_3-\alpha_1\beta_2-\alpha_2\beta_1 \\
         2\al_2\be_2-\alpha_1\beta_3-\alpha_3\beta_1 \\
    \end{array}\right)\\
\end{array}\right.
\end{array}
\]
\end{description}
The products of two irreducible representations $\bar{A}\times B$ is different from the $A\times B$ since the $T$ and the $S$ generators are complex. When we multiply the complex conjugate of a two-dimensional representation with another two-dimensional representation, we have to interchange the indices $1\leftrightarrow 2$ of the complex conjugate representation, that is, for instance in $\bar{2}\otimes 2$ we interchange $\alpha_1\leftrightarrow \alpha_2$ and the product is given by
\[
\bar{2}\otimes2=1_1\oplus1_2\oplus2 \quad
\text{with}\quad\left\{\begin{array}{l}
                    1_1\sim\alpha_2\beta_2+\alpha_1\beta_1\\[-10pt]
                    \\[8pt]
                    1_2\sim\alpha_2\beta_2-\alpha_1\beta_1\\[-10pt]
                    \\[8pt]
                    2\sim\left(\begin{array}{c}
                        \alpha_1\beta_2 \\
                        \alpha_2\beta_1 \\
                    \end{array}\right)
                    \end{array}
            \right.\\[-10pt]
\\[8pt]
\]

When we multiply the complex conjugate of a three-dimensional representation with another representation, we have to interchange the indices $2\leftrightarrow 3$ of the complex conjugate representation, that is, for instance in $\bar{3}\otimes 3$ we interchange $\alpha_2\leftrightarrow \alpha_3$. Similarly in the case of $\bar{2}\otimes 3$ we have to interchange the indices $1\leftrightarrow 2$ for the doublet, that is $\alpha_1\leftrightarrow \alpha_2$.

\end{document}